\documentclass[letterpaper,prl,twocolumn,amsmath,amssymb,groupedaddress,bibliography]{revtex4-1}
\usepackage[utf8]{inputenc}
\usepackage[draft]{hyperref}
\usepackage{graphicx}% Include figure files
\usepackage{dcolumn}% Align table columns on decimal point
\usepackage{bm}
\usepackage{amssymb}
\usepackage{physics}
\usepackage{color}
\usepackage{soul}
\usepackage{epstopdf}
\usepackage{float}

\begin{document}

\title {Polarization-Controlled Cavity Input-Output Relations }% Force line breaks with \\
%\thanks{A footnote to the article title}%
\author{Fuchuan Lei$^1$}
\email{fuchuan.lei@oist.jp} 
\author{Jonathan M. Ward$^1$}
\author{Priscila Romagnoli$^1$}
\author{ S\'ile Nic Chormaic$^{1,2}$}
\email{sile.nicchormaic@oist.jp} 

\affiliation{$^1$Light-Matter Interactions for Quantum Technologies Unit, Okinawa Institute of Science and Technology Graduate University, Onna, Okinawa 904-0495, Japan\\}
\affiliation{$^2$ Universit\'e Grenoble-Alpes, CNRS, Grenoble INP, Institut N\'eel, 38000 Grenoble, France\\}

\date{\today}% It is always \today, today,
             %  but any date may be explicitly specified

\begin{abstract}
Cavity input-output relations (CIORs) describe a universal formalism relating each of the far-field amplitudes outside the cavity to the internal cavity fields.  Conventionally, they are derived based on a weak-scattering approximation. In this context, the amplitude of the off-resonant field remains nearly unaffected by the cavity, with the high coupling efficiency into cavity modes being attributed to destructive interference between the  transmitted (or reflected) field and the output field from the cavity. In this Letter, we show that, in a whispering gallery resonator-waveguide coupled system, in the  strong-scattering regime, the off-resonant field approaches to zero, but  more than 90\% coupling efficiency  can still be achieved due to the  Purcell-enhanced channeling. As a result, the CIORs turn out to be essentially different  than in the weak-scattering regime. With this fact, we propose that the CIOR can be tailored by controlling the scattering strength. This is experimentally demonstrated 
%in a {\color{blue}{hollow whispering gallery resonator-nano fiber coupled system}} %whispering gallery resonator-waveguide coupled system,
by the transmission spectra exhibiting either  bandstop or  bandpass-type behavior according to the polarization of the input light field.

\end{abstract}

\pacs{Valid PACS appear here}% PACS, the Physics and Astronomy
                             % Classification Scheme.
%\keywords{Suggested keywords}%Use showkeys class option if keyword
                              %display desired
\maketitle

%\tableofcontents

The coupling of light from a waveguide to a microresonator can be physically treated as the scattering of a traveling wave 
by discrete localized states in the resonator \cite{fan1999theoretical,xu2000scattering}. Traditionally, the scattering is assumed to be weak and the 
coupling is characterized by almost 100\% transmission for off-resonant light. 
Interestingly, in this {\it weak-scattering} regime, a near-unity coupling efficiency can be attained if the intrinsic loss of the resonator is equal to the coupling loss induced by the waveguide; this is termed  {\it critical} coupling \cite{gorodetsky1999optical,yariv2000universal,cai2000observation}. Intrinsically, critical coupling can be regarded as a consequence of the cavity input-out relation (CIOR) in the weak-scattering regime, i.e., the perfect destructive interference between the direct transmission through the waveguide and the outcoming field from the cavity mode. Critical coupling can be considered as an example of {\it coherent perfect absorption}, which was developed in recent years \cite{chong2010coherent,wan2011time}. It has been shown that multiple critical couplings could exist within a coupled system \cite{ghulinyan2013oscillatory,acharyya2017multiple}; however, the rigorous condition required to achieve  critical coupling cannot always be satisfied. For instance, in nonlinear optics experiments \cite{del2007optical,zhang2019symmetry,carmon2007visible}, it is challenging to realize critical coupling for  two different wavebands simultaneously.

Aside from trapping  light in the cavity by creating perfect destructive interference at the coupling point, coupling of light into the cavity modes may be achieved by another mechanism - Purcell-enhanced Rayleigh scattering \cite{kippenberg2009purcell, motsch2010cavity}. This method is limited by the small scattering cross-section of the point-like (dipole) scatterers, thus it is unlikely to achieve a high coupling efficiency from the input field  \cite{zhu2014interfacing,ward2019excitation}. In fact, the Purcell effect does not necessarily have to be explained in terms of the optical density of states, but rather can be described as the  constructive interference of waves \cite{dowling1991radiation, motsch2010cavity,rybin2016purcell}. Therefore, the Purcell effect is not  purely restricted to the coupling between Rayleigh scatterers or single quantum emitters and a cavity. 
 In this Letter, we show that, when the resonator-waveguide coupled system is in the {\it strong-scattering} regime, the optical field can be strongly scattered by the resonator and the off-resonant light transmission can drop to zero;  however, the  resonant light can couple into the cavity modes with near-unity efficiency, leading to a {\it bandpass}-type transmission spectrum.  Note that we observed this effect before, but the mechanism was not  explicitly presented \cite{lei_photRes_2017}. To explain our observations, the standard input-output theory cannot be used as it is simply invalid. We provide a new theoretical interpretation by generalizing the classical Purcell effect from the dipole system to the waveguide case  and demonstrate that this phenomenon leads to near-unity efficiency coupling of light from the waveguide to the resonator. This newly discovered mechanism implies that studies of microresonators can be readily extended from the conventional weak-scattering regime to the strong-scattering regime. Based on this fact, we further propose and realize a tunable  CIOR in a resonator-waveguide coupled system. This is achieved by properly designing the geometry of the resonator-waveguide coupled system and ensuring that the coupling can be switched between the weak-scattering regime and the strong-scattering regime by simply controlling the inner degree of light, i.e., the light's polarization.

\begin{figure}[t]
\centering
\includegraphics[width=1\linewidth]{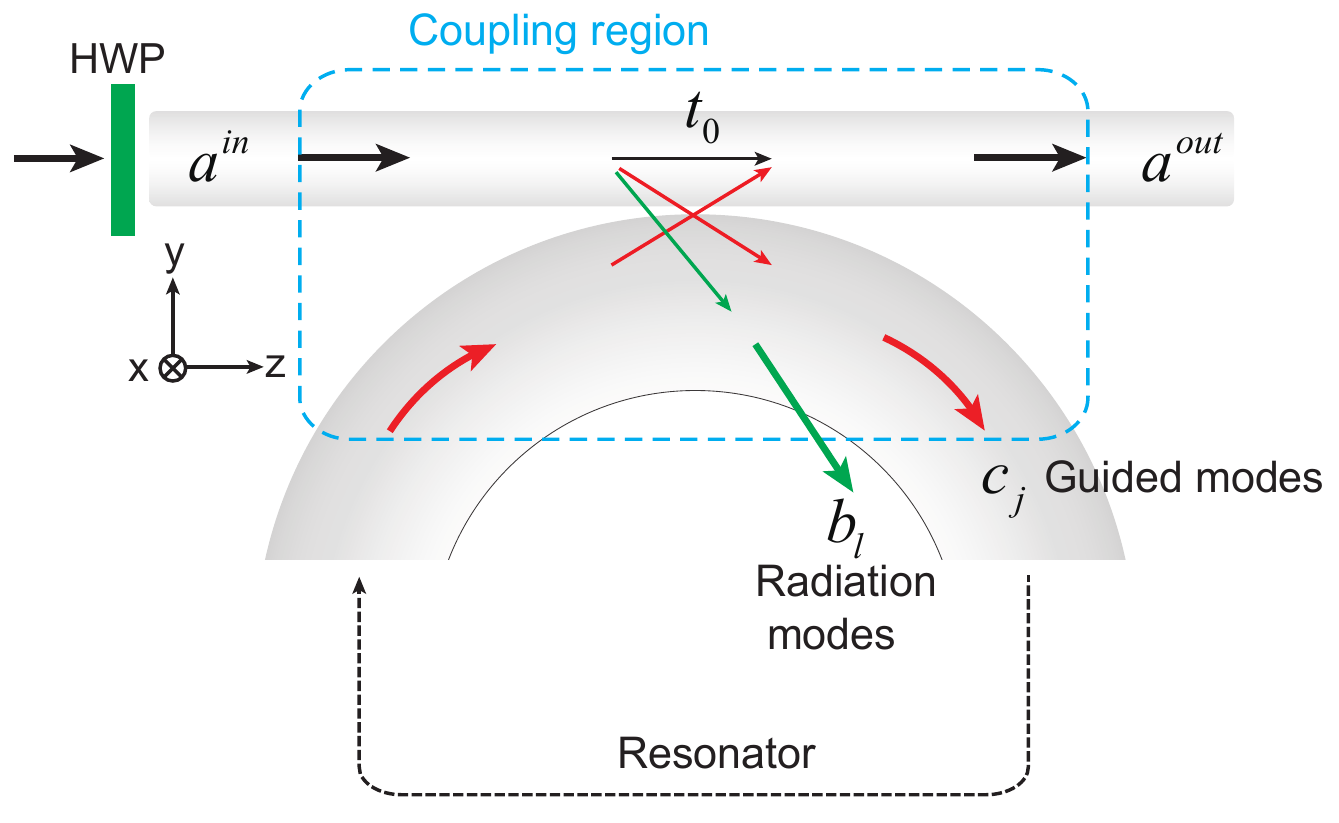}
\caption{ Schematic diagram of a single-mode fiber coupled to a whispering gallery resonator. HWP: half-wave plate. }
\label{fig:fig1}
\end{figure}

We consider a silica whispering gallery (WG) resonator coupled with an air-clad, single-mode, tapered optical fiber, as illustrated in Fig. \ref{fig:fig1}. This system has been studied extensively for more than two decades for a variety of applications  \cite{knight_ol_1997,cai2000observation,Spillane_prl_2003,matsko2006optical}.
We can treat the WG resonator as a system with two parts: (\uppercase\expandafter{\romannumeral1}) the section of the resonator in the coupling region (indicated by the dashed rectangle) and (\uppercase\expandafter{\romannumeral2}) the rest of the resonator \cite{yariv2000universal}.
In the weak-scattering regime, only one cavity mode is considered and the coupling region can be modeled as a two-port beam-splitter, i.e., in the absence of the rest of the resonator the system acts like a directional coupler having \textit{sogenannten} cavity-free guided modes. In contrast, in the strong-scattering regime (i.e., $|t_0| \approx 0$), the light can be partially coupled from the fiber mode, $a^{in}$,  not only into the guided modes, $c_{j}$, but also into a continuum of radiation modes, $b_{l}$, where $j$ and $l$ are the mode order numbers, see Fig. \ref{fig:fig1}. Strictly speaking, the guided modes should be termed as quasi-modes because they also have nonzero radiation loss. 

To illustrate the effect of the cavity on the light coupling in the strong-scattering regime, we can directly compare the coupling power into the cavity-modified guided modes, $P^c_j$,
(i.e., in the presence of part \uppercase\expandafter{\romannumeral2}) and the cavity-free guided modes, $P^w_j$. Their ratio is defined as the cavity impact factor, $G_j(\omega)$, such that \cite{supplementary} 

\begin{equation}
    G_j(\omega)=\frac{P^c_j}{P^w_j}=\frac{2\kappa_j}{[\kappa_j^2+(\omega-\omega_j)^2]\tau_j},
    \label{eq1}
\end{equation}
where $\kappa_j=\kappa^0_j+\kappa^e_j$, $\kappa^0_j$ ($\kappa^e_j$) represents the field amplitude decay rate due to the intrinsic loss (waveguide coupling).  $\tau_j$ is the circulation time for the mode  traveling inside the resonator. $\omega_j$ is the cavity resonant frequency. For the resonant case, $G_j(\omega_j)=2/ \pi(\lambda/n_j)( Q_j/L)$. Here, $n_j$ is the effective refractive index and $Q_j$ is the quality factor of the cavity mode, $j$. $L$ is the circumference of the resonator. Note that the cavity impact factor, $G$, is very similar to the well-known Purcell factor, $F ={3}/{4\pi^2} ({\lambda}/{n})^3(Q/V)$ \cite{kippenberg2009purcell}, which is widely used for dipole emitters or scatters. Essentially, the similarity between $G$ and $F$ stems from the identical underlying physics --- wave interference. Therefore, we could treat $G$ as a generalized Purcell factor. It is worth emphasizing that this cavity impact factor is invalid in the weak scattering regime therefore critical coupling cannot be attributed to the Purcell effect. Compared to the guided modes, the presence of the cavity has a much weaker effect on the distribution of the radiation modes \cite{le2009cavity}, thus we could assume the scattering rates into the radiation modes remains the same with or without the existence of part \uppercase\expandafter{\romannumeral2} of the resonator. Therefore, in the strong scattering regime, we can define the channeling efficiency, $\Gamma_j(\omega)$, which represents the fraction of power coupled from the waveguide into the cavity-modified mode, $j$ \cite{supplementary}:

\begin{equation}
    \Gamma_j(\omega)=\frac{G_j(\omega)\gamma^g_j}{\sum_k G_k(\omega)\gamma^g_k+\gamma^{rad}},
    \label{eq2}
\end{equation}
where $\gamma^g_k$ and $\gamma^{rad}$
stand for the scattering rates into the cavity-free guided mode $k$ and all radiation modes. One can see that, even if $\gamma^g_k\ll \gamma^{rad}$, the large cavity impact factor, $G_j(\omega_j)$,  can cause the channeling efficiency, $\Gamma_j(\omega_j)$, to approach to unity, e.g., $G\simeq 10^4$ for a mode with $Q = 10^7$ and $L= 400$ $\mu$m.

\begin{figure}[b]
\centering
\includegraphics[width=1\linewidth]{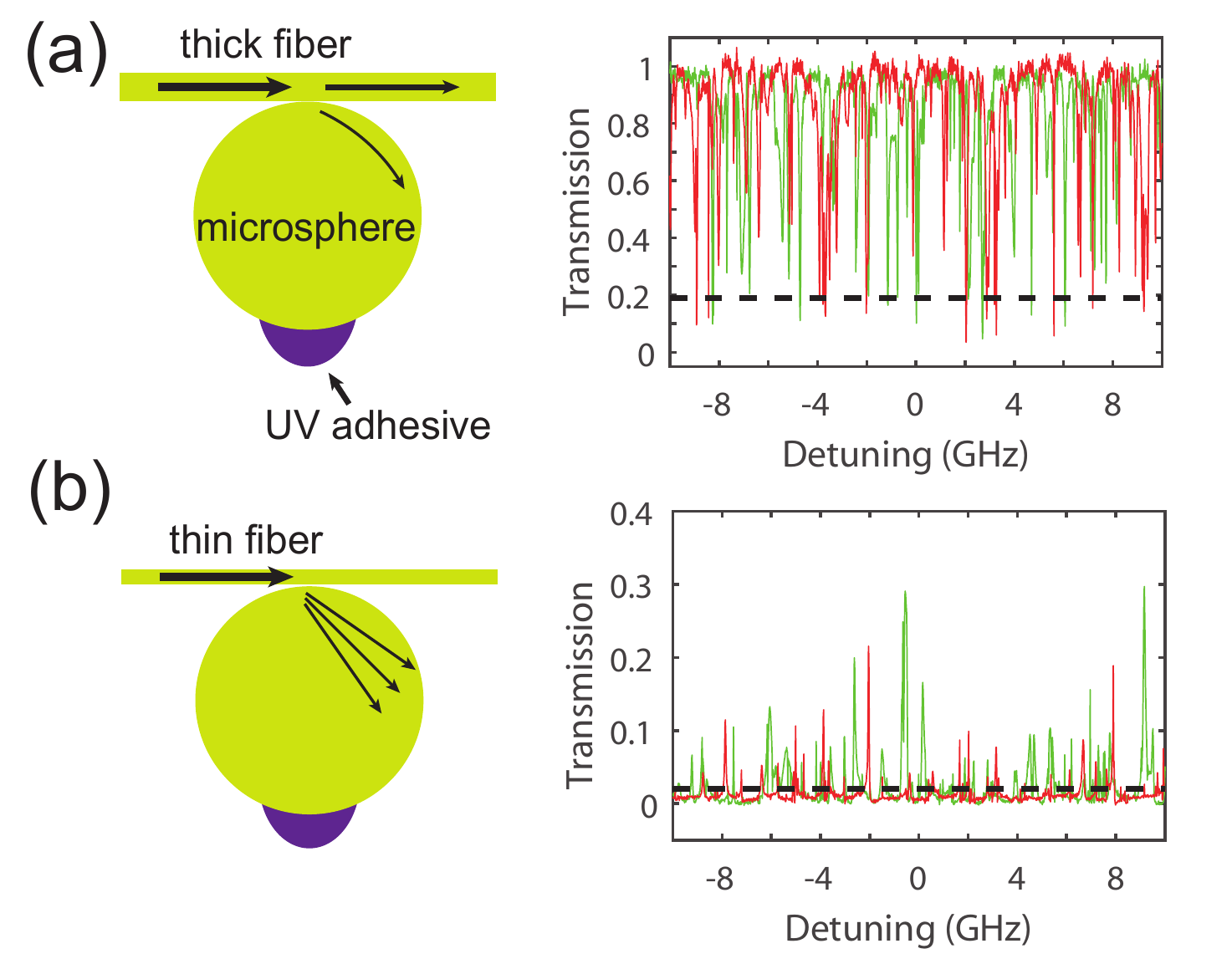}
\caption{ Study of the fiber-microsphere resonator system. The microsphere diameter is 130 $\mu$m.  The fiber diameters are (a) $d= 0.9    $ $\mu$m and (b) $d= 0.4 $ $\mu$m. The laser wavelength in this work is around 980 nm. Right-hand side: Measured transmission of the fiber-microsphere resonator system before (solid) and after coating UV adhesive (dashed) on the microsphere. The solid red and green lines correspond to an input field with horizontal ($\bf H$) and vertical ($\bf V$) polarization, respectively.}
\label{fig:fig2}
\end{figure}

In order to compare the weak-scattering and  strong-scattering regimes, we  performed measurements using a silica microsphere coupled to tapered optical fibers of different diameters  \cite{lei_photRes_2017}, as presented in Fig. \ref{fig:fig2}.
With a thick fiber, the coupling can be classified as weak-scattering: only the resonant fields are absorbed by the resonator and the transmission spectra are {\it bandstop} type, see Fig. 2(a). In contrast, with a thin fiber,  the coupling enters
the strong-scattering regime: only the resonant fields can pass through and the resulting transmission spectra are {\it bandpass} type, see Fig. 2(b). The existence of two distinct spectra is due to their distinct coupling regions. To study the coupling region  experimentally, we placed a tiny droplet of ultraviolet (UV) adhesive (NOA 81, Thorlabs) onto a small  area of the microsphere opposite to the coupling region and cured it using a UV gun, see Fig. \ref{fig:fig2}.  The adhesive has no direct influence on the coupling region, but it does act to prevent the circulation of the guided cavity modes; in other words, the WG cavity modes degrade into cavity-free guided modes. However, the radiation modes should remain nearly unmodified.

For each case, the transmission spectra are plotted for the input fields with two orthogonal polarizations, i.e., horizontal, $\bf H$ (red), and vertical, $\bf V$ (green), corresponding to the $\rm HE^x_{11}$  and $\rm HE^y_{11}$ modes in the fiber. After the adhesive was added all sharp features in the transmitted spectra disappeared and the transmission became constant  (almost identical for both $\bf H$  and $\bf V$), represented by the dashed lines in Fig. \ref{fig:fig2}. The polarization of the guided mode in the tapered fiber was controlled using the method presented in \cite{PhysRevApplied.Lei}.  During  experiments, the fiber was in contact with the resonator at all times.  

Note that, for the thicker fiber, see Fig. \ref{fig:fig2}(a), the transmissions for resonant and off-resonant frequencies are affected by the UV adhesive. 
The importance of this behavior must be stressed at this point, bearing in mind that the adhesive does not physically affect the coupling junction. For light in the fiber, the coupling system essentially looks like a directional coupler, i.e., there are no resonances and the cavity simply acts like another waveguide. On removing the adhesive,  off-resonant light cannot enter the restored cavity.

  Here, it is appropriate to point out that, for a multimode resonator, even in the weak-scattering case, the magnitude of the direct transmission coefficient, $t_0$, can be much less than unity (here $|t_0|^2 \approx 0.2$). This is not in conflict with the widely used assumption in this regime, i.e., that the off-resonant field passes through the coupling region with unity transmission, since the real direct transmission coefficient, $t_j$, is actually modified by all cavity modes. As a result, for the weak-scattering case, $t_j=1 \neq t_0$, and the CIOR of a multimode resonator in the vicinity of $\omega_j$ is the same as for a single-mode resonator  \cite{haus1984waves,gardiner1985input}:    
\begin{equation}
    a_j^{out}=a_j^{in}-\sqrt{2\kappa^e_j}c_j.
    \label{eq3}
\end{equation}

For a thin fiber, as shown in Fig. \ref{fig:fig2}(b), the light is strongly scattered so it cannot pass through the coupling region without the assistance of the cavity modes. The non-vanishing (larger than 20$\%$ for some modes) transmission of resonant light originates from the Purcell-enhanced channeling. Thus,  the CIOR in the strong-scattering regime is \cite{supplementary}:
\begin{equation}
    a_j^{out}=\sqrt{\frac{\Gamma_j(\omega)\kappa^e_j}{\kappa^e_j+\kappa^0_j}}a_j^{in}.
    \label{eq4}
\end{equation}

\begin{figure}[b]
\centering
\includegraphics[width=0.95\linewidth]{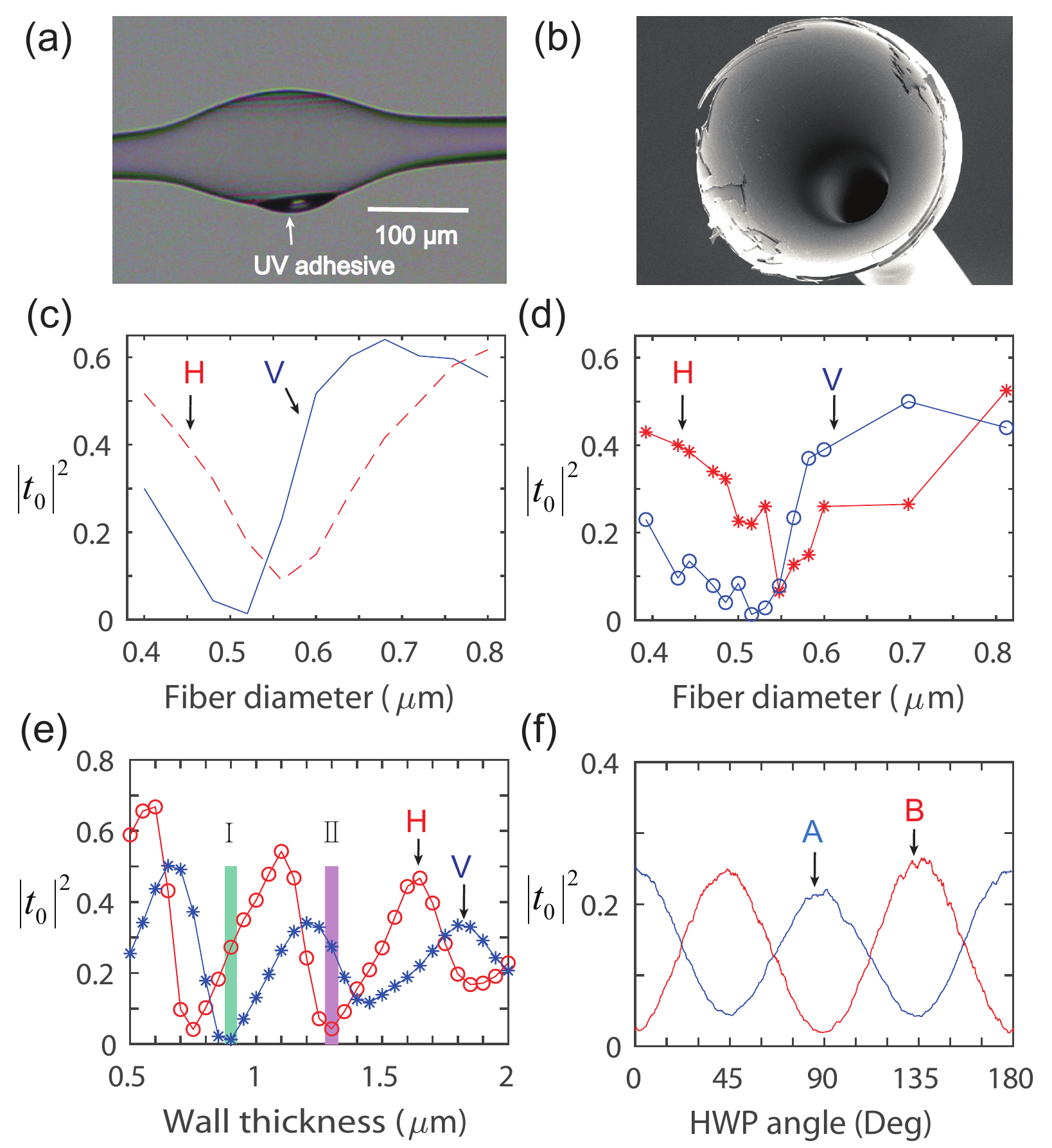}
\caption{{ Study of the fiber-microbubble resonator system.} (a) A silica microbubble resonator with UV  adhesive on one side. (b) Scanning electron microscope (SEM) image of the cross-section of the microbubble.  (c) 
Finite element method (FEM) calculation of the fiber transmission with ${\bf H}$ and ${\bf V}$ polarizations through the coupling region (the blue dashed box in Fig. \ref{fig:fig1}). We used a microbubble diameter of 120 $\mu$m and a wall thickness of $w=0.9 $ $\mu$m. (d) Measured transmission through a tapered fiber of different diameters coupled to a microbubble with UV adhesive present.  The small fluctuations could be caused by a change to the coupling position when translating the tapered fiber. (e) Same as (c) except for different wall thicknesses and a fiber diameter of $d=0.5$  $\mu$m. (f)
Measured transmission of two fiber-microbubble (coated with UV adhesive) coupling systems (Samples A and B) for different input polarizations.  The HWP angle $0^{\circ}$ ($45^{\circ}$) corresponds to $\bf {H}$ ($\bf V$).  Here, the fiber diameter $d= 0.5$ $\mu$m and $w= 0.9\pm0.05 $ $\mu$m for sample A (blue),  $w =1.3\pm0.05 $ $\mu$m for sample B (red).}
\label{fig:fig3}
\end{figure}

 When the scattering strength lies between the weak and  strong regimes, the output field is the superposition of the partial directly transmitted field  and the field leaking out from the cavity mode. Therefore, the coupling mechanism cannot be simply attributed to the destructive interference-induced trapping  or Purcell-enhanced channeling.  In this regime, a high coupling efficiency is unlikely to be achievable, since there needs to be a balance between the directly transmitted, intracavity, and radiation fields.

The existence of radically different CIORs for the weak and scattering coupling regimes implies that the resonances can be selectively controlled to induce either absorption or transparency of light.
However, in general, for  a given set of system parameters, e.g., the fiber size used above, it is difficult to utilize both the bandpass and bandstop functions. By noting the CIOR is determined from the scattering strength (or $t_0$) it is feasible to achieve a tunable CIOR in a system with a  polarization-dependent scattering strength (or $t_0$).

Here, we achieve this goal using a hollow microbubble WG resonator, as shown in Fig. \ref{fig:fig3}(a). The hollow microbubble was fabricated using a $\rm CO_2$ laser focused onto a silica microcapillary \cite{watkins2011single} and the wall of the microbubble can be as thin as a few hundred nanometers \cite{yang2016high}.
When a tapered fiber couples to the  resonator, the coupling region of the system may demonstrate strong birefringence due to its unique geometry. Specifically, the thin wall of the resonator may act as a curved 2D waveguide that can support two polarized, guided modes, i.e., TE and TM modes with different propagation constants, in addition to radiation modes. Therefore, the coupling coefficients, such as $t_0$, are quite sensitive to the polarization of the input field.  A calculation of $|t_0|^2$ through the coupling region was performed using a finite element method (COMSOL), the results of which are shown in Figs. \ref{fig:fig3}(c) and (e). In contrast to solid microsphere resonators  \cite{lei_photRes_2017},  ${|t_0|}^2$ does not monotonically depend on the fiber diameter, $d$, and the resonator wall thickness, $w$. The phenomenon is  reminiscent of two coupled waveguides, where the optical energy is periodically exchanged between them.

Using the same technique as before (i.e., transmission measurements with a  fiber and UV adhesive on the resonator), $|t_0|^2$ for the microbubble-fiber coupled system can be measured experimentally. The measured transmittance as a function of fiber diameter is shown in Fig. \ref{fig:fig3}(d) and the measured data correspond well with the simulated results plotted in Fig. \ref{fig:fig3}(c).   For some specific sets of parameters, e.g., $w=0.9$ $\mu$m and $d=0.5$ $\mu$m (color region \uppercase\expandafter{\romannumeral1}), and $\bf H$ polarization at the input,  reasonable transmission is observed ($|t_0|^2=0.3$), whereas for $\bf V$ input polarization it is almost completely lost ($|t_0|^2 \approx 0$). Therefore, unlike for microspheres,   polarization-controlled CIORs should be achievable in this system. Certainly, in  resonators with different geometries, it is possible to have  CIORs with different dependence on the polarization, see the color regions in Fig. \ref{fig:fig3}(e).  Figure \ref{fig:fig3}(f) shows how the measured  ${|t_0|}^2$, varies as a function of the input polarization for two different microbubbles with wall thicknesses corresponding to the color regions, \uppercase\expandafter{\romannumeral1} and \uppercase\expandafter{\romannumeral2}, in Fig. \ref{fig:fig3}(e). 

\begin{figure}[t]
\centering
\includegraphics[width=0.9\linewidth]{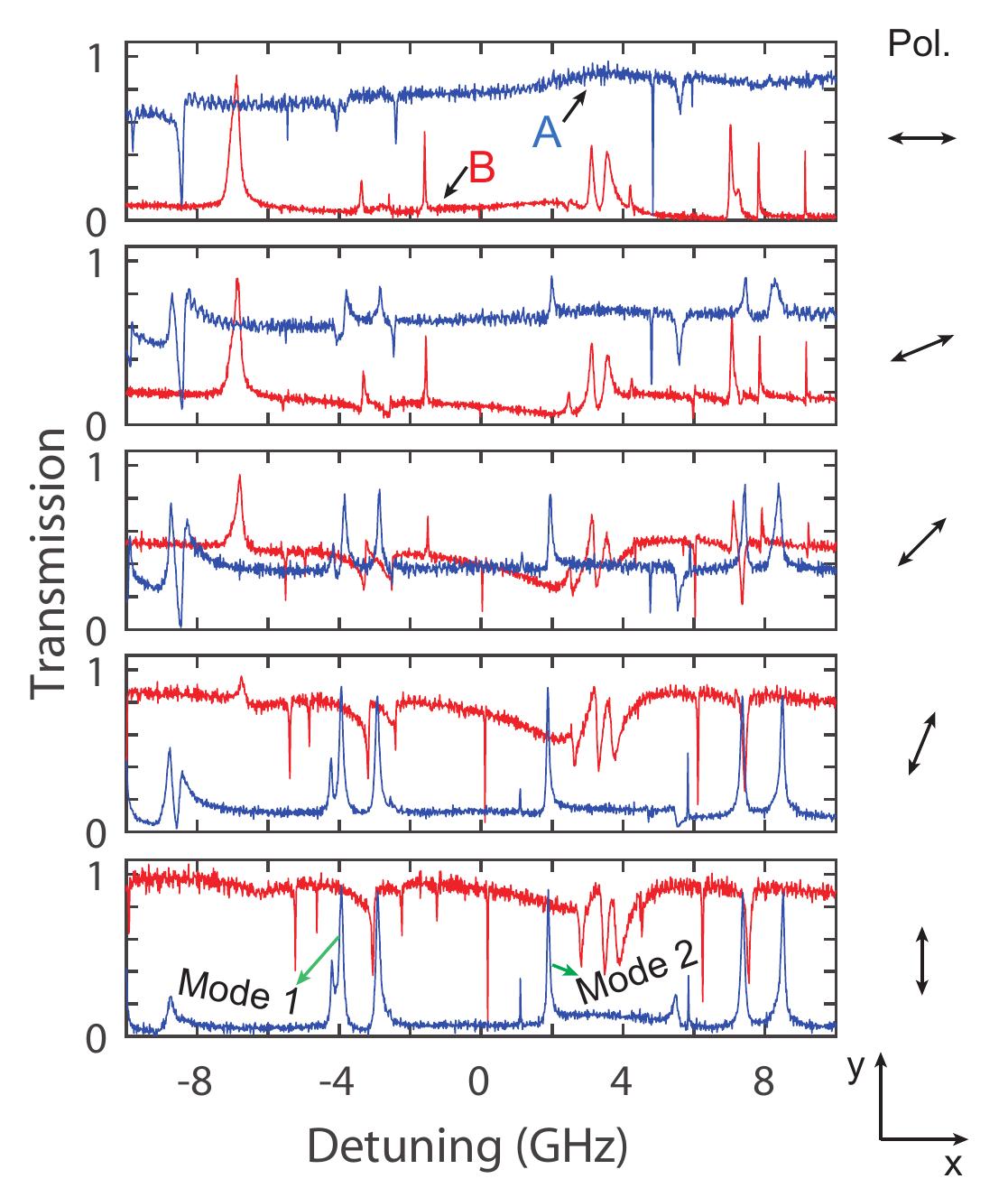}
\caption{Experimental demonstration of polarization-controlled CIOR with two microbubble resonators, samples A (blue) and B (red), corresponding to Fig. \ref{fig:fig3}(f). The transmittance is normalized to the transmission of the bare fiber. From top to bottom the input polarization changes from $\bf H$ to $\bf V$.}
\label{fig:fig4}
\end{figure}

To demonstrate the feasibility of achieving a polarization-controlled CIOR, the transmission spectra of two microbubble samples are measured and presented in Fig. \ref{fig:fig4}.  With the input polarization changing from $\bf H$ to $\bf V$, the transmission spectrum evolves continuously from a {\it bandstop} to a {\it bandpass} type for sample A (blue) and the opposite for sample B (red). During this process, the excited cavity modes also switch from TE to TM.

Comparing the results obtained using a microbubble to those for a microsphere resonator (Fig. \ref{fig:fig2}), the highest peaks in the transmission spectra in the  strong-scattering regime
are much higher, with the measured maximum value exceeding 93$\%$ (Mode 1 in Fig. \ref{fig:fig4}).  This can be understood based on Eq. (\ref{eq4}).
The higher transmission can be achieved at the expense of increasing the coupling rates, i.e., decreasing the external $Q$-factor. For instance, with an intrinsic $Q_0=10^8$ and external $Q_e=5 \times 10^6$, $T(\omega_j)=0.95 \times \Gamma_j$.  Due to the unique geometry of the microbubble resonator, the effective refractive indices of the cavity modes are closer to that of the fiber mode compared to those of a microsphere. Accordingly, the relatively large coupling leads to a reduction in the total $Q$-factors of those cavity modes, see Fig. \ref{fig:fig4}. 
Nevertheless, a $Q$-factor of $10^7$ can be achieved (Mode 2 in Fig. \ref{fig:fig4}) and its transmission is 91$\%$. It is worth noting that the determination of the Q-factor in the strong-scattering case is not the same as for the weak-scattering, where one can simply consider the {FWHM} as { $\kappa_j/\pi$}. Here, { $\kappa_j/\pi$} can be obtained by fitting with Eq. (\ref{eq4}) and it is smaller than the {FWHM} of the peak in the spectrum. The observation of the high transmission peaks also demonstrates that the Purcell-enhanced channeling efficiency, $\Gamma_j$, can indeed approach  unity.  Finally, we emphasize that the
realization of polarization-controlled CIORs should be attributed to the \textit{cavity-modified} birefringence. By comparing the transmission of the off-resonant light in Fig. \ref{fig:fig3}(f) and Fig. \ref{fig:fig4}, one can see that the birefringence effect can be enhanced by the resonator.

To conclude, we have shown that, due to the existence of a Purcell-enhanced channeling mechanism,  applications of resonator-waveguide systems can be extended from the widely used weak-scattering regime to the strong-scattering regime, where a new CIOR can be exploited. By taking advantage of the fact that the scattering strength can be switched between the weak and strong regimes by changing the input polarization, we have shown that a tunable CIOR is achievable in a specific resonator-waveguide system. This counter-intuitive demonstration of a tunable CIOR could have wide impact in designing optical circuits \cite{jin2018inverse} for optical switching \cite{stegmaier2017nonvolatile}, tunable filtering \cite{rokhsari2004ultralow,lei_photRes_2017}, and integrated polarization elements \cite{crespi2011integrated,shen2015integrated}. The polarization-dependent CIOR facilitates the preparation of entangled states in a cavity quantum electrodynamics (cQED) system in a novel way \cite{dayan2008photon} and could be used for  cavity-based quantum information processing \cite{volz2014nonlinear,shomroni2014all}. Last, but not least, the Purcell-enhanced channeling could be used as a complementary method  to achieve high-efficiency coupling where  critical coupling cannot be accessed, for example, in broadband frequency comb generation \cite{del2007optical}, {or second- and third-harmonic generation \cite{zhang2019symmetry,carmon2007visible}}.

 This work was supported by 
Okinawa Institute of Science and Technology
Graduate University (OIST).

\bibliography{ref}
\pagebreak
\widetext
\begin{center}
    \textbf{\large Supplemental Materials: Polarization-controlled cavity input-output relations}
\end{center}

\section*{Theory model details}

In the presence (absence) of Part \uppercase\expandafter{\romannumeral2} of the resonator, see Fig. 1, the amplitude of a guided mode $j$ (with $1\leq j \leq N$)  is denoted as $A^c_j$ ($A^w_j$). Let $A_{in}$ be the slowly varying amplitude with $|A_{in}|^2$ equal to the total input power. We introduce the coefficients, $t_j$, to describe the coupling of the fiber mode with the guided modes such that
\begin{equation}
    A^w_j=t_jA_{in}.  \tag{S.1}\label{eq:S.1}
\end{equation}
Hence, the corresponding coupling power, $P^w_j$, is given by 
\begin{equation}
    P^w_j=|t_j|^2|A_{in}|^2. \tag{S.2}\label{eq:S.2}
\end{equation}

\noindent Next, let us take the resonances into consideration.  The equation describing a cavity mode, $j$, is \cite{gorodetsky1999optical}
\begin{equation}
    A^c_j(t)=R_jA^c_j(t-\tau_j){\rm exp}(-\alpha L/2+i2\pi n_j L/\lambda)+t_jA_{in}, \tag{S.3}\label{eq:S.3}
\end{equation}
where $\tau_j =n_j L/c$ is the circulation time for the mode traveling inside the resonator and $L$ is the circumference of the resonator. $n_j$ is the effective refractive index of the mode $j$, and $\lambda$ is the wavelength. $R_j$ stands for the internal reflectively coefficient when the mode passes through the coupling region. After some simple algebra, we arrive at

\begin{equation}
    \frac{dA^c_j}{dt}+(\kappa^e_j+\kappa^0_j+i\Delta\omega)A^c_j=\frac{t_j}{\tau_j}A_{in} \tag{S.4}\label{eq:S.4}
\end{equation}
where 
\begin{equation}
    \kappa^e_j=\frac{1-R_j}{R_j\tau_j}, \kappa^0_j=\frac{\alpha c}{2n_j}. \tag{S.5}\label{eq:S.5}
\end{equation}
For the steady-state, we have that
\begin{equation}
   A^c_j=\frac{t_jA_{in}}{(\kappa^e_j+\kappa^0_j+i\Delta\omega)\tau_j}. \tag{S.6}\label{eq:S.6}
\end{equation}
In the strong-scattering regime, the optical field does not pass through the coupling regime directly. When in the steady state, the power, $P^c_j$, coupled into the cavity mode, $j$,  equals its total loss, including intrinsic and  extrinsic losses (i.e., leakage into the waveguide).  Hence, we have 
\begin{equation}
    P^c_j=2|A^c_j|^2\tau_0(\kappa^e_j+\kappa^0_j).\tag{S.7}\label{eq:S.7}
\end{equation}
If we define the ratio of $P^c_j$ to $P^w_j$ as the cavity impact factor, $G_j(\omega)$, one finds
\begin{equation}
    G_j(\omega)=\frac{P^c_j}{P^w_j}=\frac{2\kappa_j}{[\kappa_j^2+(\omega-\omega_j)^2]\tau_j}, \tag{S.8}\label{eq:S.8}
\end{equation}
where $\kappa_j=\kappa^0_j+\kappa^e_j$.

It is worth pointing out the coupling efficiency into the cavity mode is not solely determined by the resonance itself, but also by all other radiation modes. To make it clear, one can consider that there exists a radiation source at the coupling region which is driven by the input field. Because all modes are coupled with the radiation source, the fraction of power coupled from the waveguide into the cavity-modified mode, $j$, is proportional to the coupling rate. This is  similar to the case of cavity-modified spontaneous radiation of quantum emitters \cite{le2009cavity,PhysRevB.75.205437,yalla2014cavity}.   By considering Eq. (2), it is straightforward to obtain the cavity input-output relation for the strong-scattering regime (i.e., Eq. (4)) by considering the optical energy loss through the cavity mode, with the output field arising from the extrinsic loss of the cavity mode. The absolute phase of the output field is not important and it can be ignored by properly defining $a^{out}_j$.

\end{document}